\documentclass{article}
\usepackage[utf8]{inputenc}
\usepackage{graphicx}
\usepackage{wrapfig}
\usepackage[numbers,sort&compress]{natbib}
\newcommand{\RomanNumeralCaps}[1]{\MakeUppercase{\romannumeral #1}}
\usepackage{authblk}
\usepackage[left=3cm,right=3cm, top=3cm,bottom=3cm]{geometry }
\usepackage{float}
\usepackage[symbol]{footmisc}

\newcommand{\Ca}{$^{40}$Ca$^+\;$}
\newcommand{\SDl}{P$_{1/2}\!\leftrightarrow\,$D$_{3/2}\;$}

\newcommand{\SPh}{S$_{1/2}\!\leftrightarrow\,$P$_{3/2}\;$}
\newcommand{\SPl}{S$_{1/2}\!\leftrightarrow\,$P$_{1/2}\;$}
\newcommand{\DhPh}{D$_{5/2}\!\leftrightarrow\,$P$_{3/2}\;$}
\newcommand{\DlPh}{D$_{3/2}\!\leftrightarrow\,$P$_{3/2}\;$}
\newcommand{\DlPl}{$P_{1/2}\!\leftrightarrow\,$D$_{3/2}\;$}
\newcommand{\Sl}{S$_{1/2}\;$}
\newcommand{\Pl}{P$_{1/2}\;$}

\newcommand{\Dl}{D$_{3/2}\;$}
\newcommand{\Dh}{D$_{5/2}\;$}

\begin{document}

\title{\vspace{-1cm} Enhanced ion-cavity coupling through cavity cooling in the strong coupling regime}

\author[1]{Costas Christoforou\thanks{Email: C.Christoforou@sussex.ac.uk}}
\author[1]{Corentin Pignot}
\author[2]{Ezra Kassa}
\author[3,4]{Hiroki Takahashi}
\author[1]{Matthias Keller}
\affil[1]{ Department of Physics and Astronomy, University of Sussex, Brighton, BN1 9QH, United Kingdom }
\affil[2]{ Clarendon Laboratory, University of Oxford, Parks Road, Oxford OX1 3PU, United Kingdom}
\affil[3]{Quantum Information and Quantum Biology Division,
Institute for Open and Transdisciplinary Research Initiatives,
Osaka University, 1-3 Machikaneyama, Toyonaka, Osaka 560-8531, Japan}
\affil[4]{Experimental Quantum Information Physics Unit,
Okinawa Institute of Science and Technology Graduate University,
1919-1 Tancha, Onna, Kunigami, Okinawa 904-0495, Japan}
\maketitle

\begin{abstract}

Incorporating optical cavities in ion traps is becoming increasingly important in the development of photonic quantum networks. However, the presence of the cavity can hamper efficient laser cooling of ions because of geometric constraints that the cavity imposes and an unfavourable Purcell effect that can  modify the cooling dynamics substantially. On the other hand the coupling of the ion to the cavity can also be exploited to provide a mechanism to efficiently cool the ion. In this paper we demonstrate experimentally how cavity cooling can be implemented to improve the localisation of the ion and thus its coupling to the cavity. By using cavity cooling we obtain an enhanced ion-cavity coupling of $2\pi \times (16.7\pm 0.1)$ MHz, compared with $2\pi \times (15.2\pm 0.1)$ MHz when using only Doppler cooling.
\end{abstract}

\section{Introduction}

Incorporating optical cavities into ion traps merges the outstanding properties of ions such as long coherence times \cite{PhysRevLett.113.220501} and high-fidelity quantum control \cite{PhysRevLett.117.060504} with the means to deterministically transfer the quantum state between ions and light at the single quantum level. Exploiting this superior control, cavity induced transparency \cite{PhysRevLett.124.013602}, the mapping of the quantum state between ions and photons \cite{Nature.485.7399}, ion-photon entanglement \cite{NaturePhot.7.3} and heralded ion-ion entanglement \cite{PhysRevLett.111} have been demonstrated. Ion-cavity systems enable distributed architectures for large-scale quantum information processing as well as device-independent quantum key distribution.
Instrumental to these applications is the ability to localise ions in optical cavities \citep{Nature.414.6859, PhysRevLett.89.103001, PhysRevLett.116.223001, JModPhys.65.5} and to strongly couple single ions to an optical cavity \cite{PhysRevLett.124.013602}.

While the optical cavity is crucial to facilitate the ion-light interaction, its presence can interfere with the manipulation of the ion. In particular laser cooling can be hindered by the Purcell effect \cite{PhysRevA.96.023824}. This can be circumvented by choosing cooling transitions within the ion that are not affected by the cavity \cite{PhysRevLett.124.013602}. Alternatively, the ion-cavity interaction can be exploited to improve the cooling process.
Cooling of trapped ions by using their interaction with an optical cavity has been theoretically studied extensively in the weak binding regime \citep{OptComm.97.353, PhysRevLett.79.4974, PhysRevA.58.3030, PhysRevLett.95.143001, NJP.14.023002}, where the transition linewidth is larger than the secular frequencies. Cavity sideband cooling in the strong binding regime has been studied both theoretically \cite{PhysRevA.64.033405} and experimentally \cite{PhysRevLett.103.103001, schleier2011optomechanical}. 

Here we demonstrate that cavity cooling in the weak binding regime can be employed to improve the localisation of a single ion in an optical cavity and to enhance the ion-cavity coupling. This paper is structured as follows. In section 2 we present the setup of our experiment. Section 3 describes the implementation of cavity cooling and finally section 4 presents the measurement of the improved ion-cavity coupling.

\section{Setup}
\begin{figure}[h]
    \centering
	\includegraphics[scale=0.52]{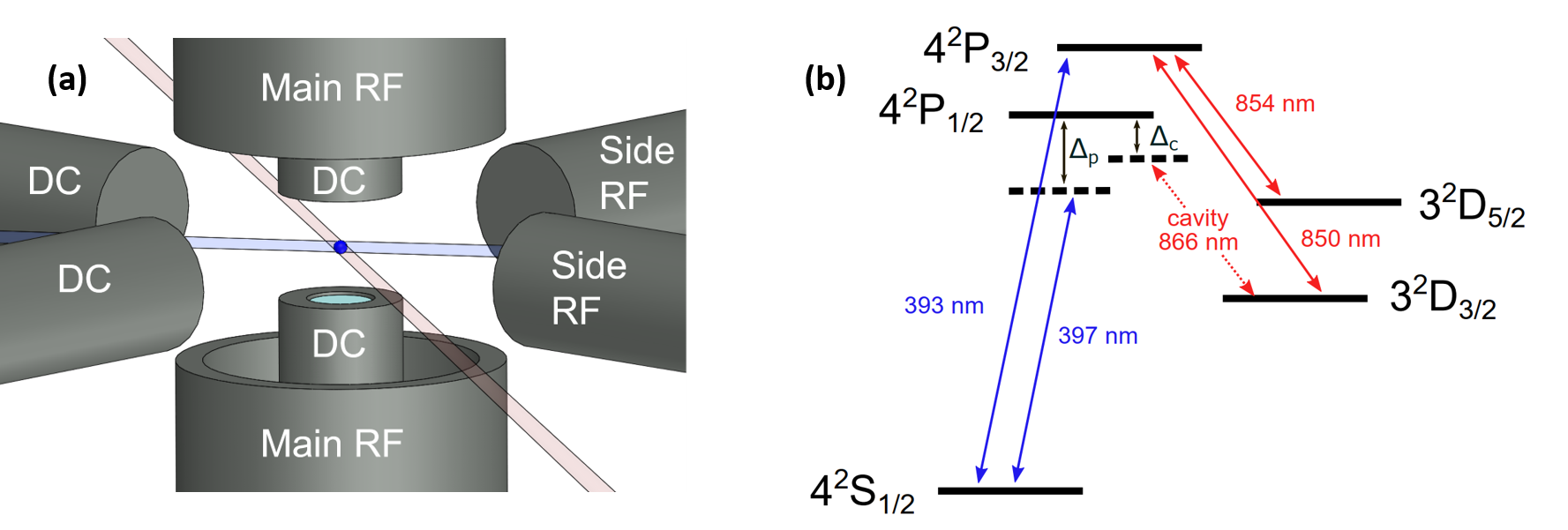}
	\caption{(a) A drawing of the ion trap. Outer axial electrodes provide the main rf signal. Two axial and two radial dc electrodes are used for micromotion compensation. Additionally, two electrodes in the radial plane carry rf signals to optimise the ion's position overlap with the cavity mode. The cavity fibres are inserted into the inner electrodes. (b) Relevant energy levels of $^{40}\mathrm{Ca}^+$ in the form of a lambda-type  system where single photons are generated at a wavelength of 866 nm via a cavity-assisted Raman transition from the $\mathrm{S}_{1/2}$ to  the $\mathrm{D}_{3/2}$ state. The ion is Doppler cooled on the \SPl and \SPh transitions. Lasers at 850 nm and 854 nm are necessary to repump from the metastable D states. The fibre cavity is tuned to the \DlPl transition.}
	\label{fig_Trap}
\end{figure}

The trap used in this experiment is an endcap-style radio frequency (rf) Paul trap with an integrated fibre Fabry-Pérot cavity (FFPC) as described in \cite{PhysRevA.96.023824}. The trapping structure consists of two electrode assemblies made from concentric cylindrical stainless steel tubes (Figure \ref{fig_Trap}(a)). An rf voltage at 19.55 MHz is applied to the outer electrodes whilst the inner electrodes are held at rf ground. This generates a trapping pseudopotential with a global minimum between the two electrode assemblies. In addition, the trap has four radial electrodes as shown in Figure \ref{fig_Trap}(a). Two of them deliver dc voltages to compensate excess micromotion in the radial plane whilst the other two radial electrodes carry rf voltages synchronous with the trap drive to move the potential minimum and optimize the ion-cavity coupling \cite{JModPhys.65.5}. The axial micromotion component is compensated by dc voltages applied to the inner electrodes of the endcap trap. The axial secular frequency for these measurements is measured to be $2\pi\times2.73$ MHz.

The fibres that carry the cavity mirrors are inserted into the inner electrodes. The end facets of the fibres were machined using a laser ablation technique \cite{Takahashi:14} and coated with high reflective coatings. An FFPC is then formed along the axis of the trap. The input of the cavity is a single-mode fibre whilst the output is a multi-mode fibre for high collection efficiency of the cavity emission. The length of the cavity is stabilised using the Pound-Drever-Hall technique with a laser at a wavelength of 897 nm, which does not interact with the ion. This laser is injected into the cavity through the single-mode input fiber and is, in turn, stabilised to a reference laser via a transfer cavity \cite{RevSciInstr.81.075109}. The cavity length is 370 $\mathrm{\mu m}$, providing a small cavity mode volume, with a theoretical maximum ion-cavity coupling of $2\pi\times$ 17.3 MHz. 
We measure a cavity linewidth at the ion's transition wavelength at 866 nm of $2\pi\times (8.2 \pm 0.1)$ MHz and a cavity finesse of 50000.

Figure \ref{fig_Trap} (b) shows the relevant energy levels of the \Ca ion used in the experiment. The ion is Doppler cooled on the \SPl and S$_{1/2}\leftrightarrow$P$_{3/2}$ transitions and the population in the \Dl and \Dh states is repumped with lasers resonant with the \DlPh and \DhPh transitions. The cavity length is tuned such that the cavity frequency is  detuned by $\Delta_c$ from the \DlPl transition. Single photons are generated by a cavity-assisted Raman transition from the \Sl to the \Dl state. The decay rate of the \Pl state is $\gamma = 2 \pi \times 11.5$ MHz. 

In order to optimise the ion's axial overlap with the standing wave pattern of the longitudinal cavity mode, the ion is translated by applying a small dc voltage on the inner axial electrodes. Although applying an rf-potential to the inner electrodes instead would be preferable \cite{JModPhys.65.5}, it is difficult to rapidly modulate the axial rf voltages on the inner electrodes while simultaneously taking into account their interference with the rf potentials on the radial electrodes due to their mutual capacitive couplings. The ion is translated by applying a positive voltage to one electrode and the corresponding negative voltage to the opposite electrode.

\begin{figure}[h]
    \centering
	\includegraphics[scale=0.65]{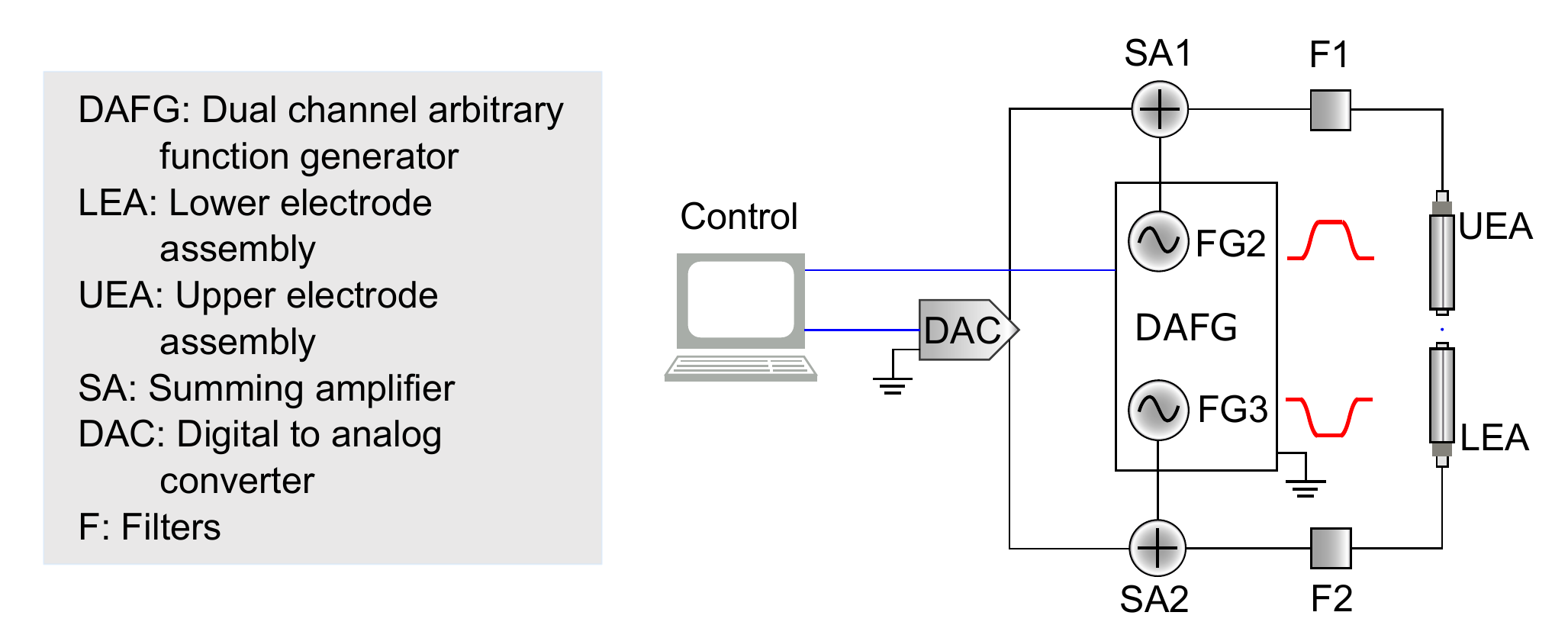}
	\caption{Schematic of the electrical setup. A DAFG provides the necessary voltages to shuttle the ion from the node to the antinode. These voltages are summed to dc voltages provided by a DAC used for the  excess micromotion compensation. The total voltages are applied to the inner electrodes, shuttling the ion along the cavity axis.}
	\label{fig_ElectricalSetup}
\end{figure}

While the coherent coupling is optimal at the antinode of the cavity field, the emission of the ion into the mode results in heating of the ion's motion along the cavity axis. For optimal axial cavity cooling, the ion's position must be moved to the node of the cavity field \cite{maunz2004cavity}. To probe the ion when it is maximally coupled to the cavity while still benefiting from efficient cavity cooling, the ion must be moved from the node, where it is cooled, to the antinode where it is probed. For this purpose, a dual channel arbitrary function generator (DAFG) provides voltages to the upper and lower electrodes to shuttle the ion. In this case we avoid introducing a static quadrupole by applying the same signal to the two inner electrodes but with an opposite sign. The ion is transferred adiabatically in order to avoid motional excitation. To this end, the signal applied to the rf ground electrodes follows a Blackman-Nuttal window (BNW) shape, the amplitude of which can be tuned to precisely shuttle the ion from a node to any desired point in the standing wave (Figure \ref{fig_ElectricalSetup}). The shuttling voltage is summed with the dc micromotion compensation voltage and applied to the inner electrodes. Low pass filters are used to reduce electrical noise from the summing amplifier and block the rf signal coming from the trap. An example of the shape of the shuttling signal can be seen in Figure \ref{fig_pulse}. \newline

\section{Cavity Cooling Implementation}
 For cavity cooling, a laser on the \SPl transition (pump laser) is tuned close to Raman resonance with the cavity on the \SDl transition. Simultaneously applying lasers on the \DlPh and \DhPh transitions provides re-pumping of the ion back into the \Sl state, completing the cycle for continuous emission of photons into the cavity. The laser on the \SPl transition is red-detuned with respect to the atomic resonance and hence provides Doppler cooling. However, due to the almost perpendicular alignment of the laser with respect to the cavity axis, Doppler cooling along the cavity axis is inefficient. 
 
  The cavity, together with the pump laser, forms a Raman transition between the \Sl and the \Dl states. When the ion is moved to a node of the cavity field, this Raman transition couples not only the internal state of the ion but also to the ion's axial motion \cite{maunz2004cavity}. Tuning the cavity blue with respect to the Raman resonance leads to a net reduction of the motional excitation of the ion, similar to Raman sideband cooling. Hence, the emission of a cavity photon leads to a reduction in the motional state excitation of the ion \cite{OptComm.97.353}. In order to measure the improvement of the ion's localisation due to the cavity cooling process we use the standing wave pattern measured by the cavity emission from the ion, similarly to  \cite{Nature.414.6859}. Increase in the visibility indicates an increase in the ion's localisation.
  
\begin{figure}[H]
    \centering
	\includegraphics[scale=0.60]{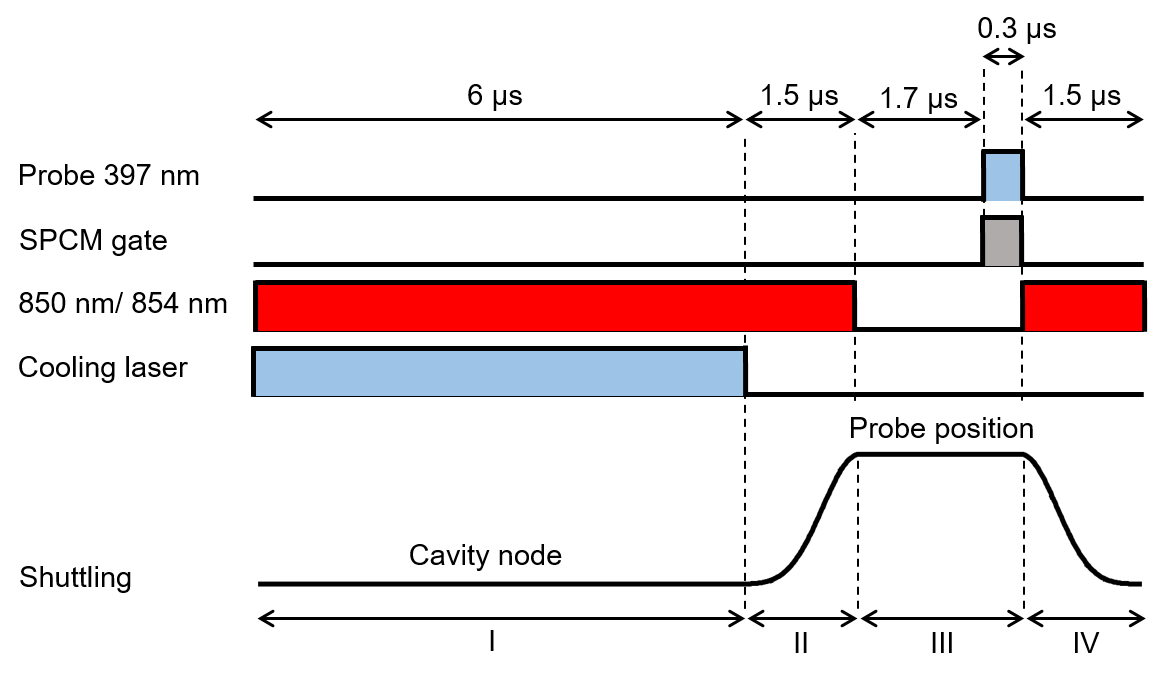}
	\caption{Laser pulses and shuttling voltage pulse of the experimental sequence. (I) Cavity cooling of the ion at the node. (II) Adiabatic shuttling of the ion to the new position and state preparation to the $4^{2}\mathrm{S}_{1/2}$ state. (III) Single photon emission in the cavity by driving the Raman transition. Photons are detected by a single-photon counting module (SPCM) connected to the output of the fibre cavity. (IV) Adiabatic return of the ion to the node. }
	\label{fig_pulse}
\end{figure}

To map the standing wave pattern the following experimental sequence is applied (Figure \ref{fig_pulse}). The ion is initially placed at the node of the cavity mode and the lasers on the \SPl, \DlPh and \DhPh transitions are applied. The cavity is blue-detuned relative to the nominal Raman resonance to facilitate cavity cooling. After the cooling period the laser on the \SPl transition is extinguished to prepare the ion in the \Sl state, and then the ion is adiabatically shuttled to the new position. At the new position a probe laser produces a single photon in the cavity using a cavity-assisted Raman resonance. The cavity emission is recorded with a single-photon detector that is gated during the probe laser pulse. There is a 1.7 $\mu$s delay before the single photon detection in order to eliminate the effect of transient voltages in the shuttling process. The ion is probed for 300 ns. The sequence is concluded by adiabatically shuttling the ion back to the node of the cavity field while repumping with the lasers on the \DlPh and \DhPh transitions. This sequence operates at a rate of 91 kHz. Scanning the final position of the ion after cooling, by varying the amplitude of the shuttling pulse, provides the position-dependent photon emission probability. The sequence was repeated 180000 times at each final position and the errors were extracted from the variations between the measurements.

\begin{figure}[h]
    \centering
	\includegraphics[scale=0.30]{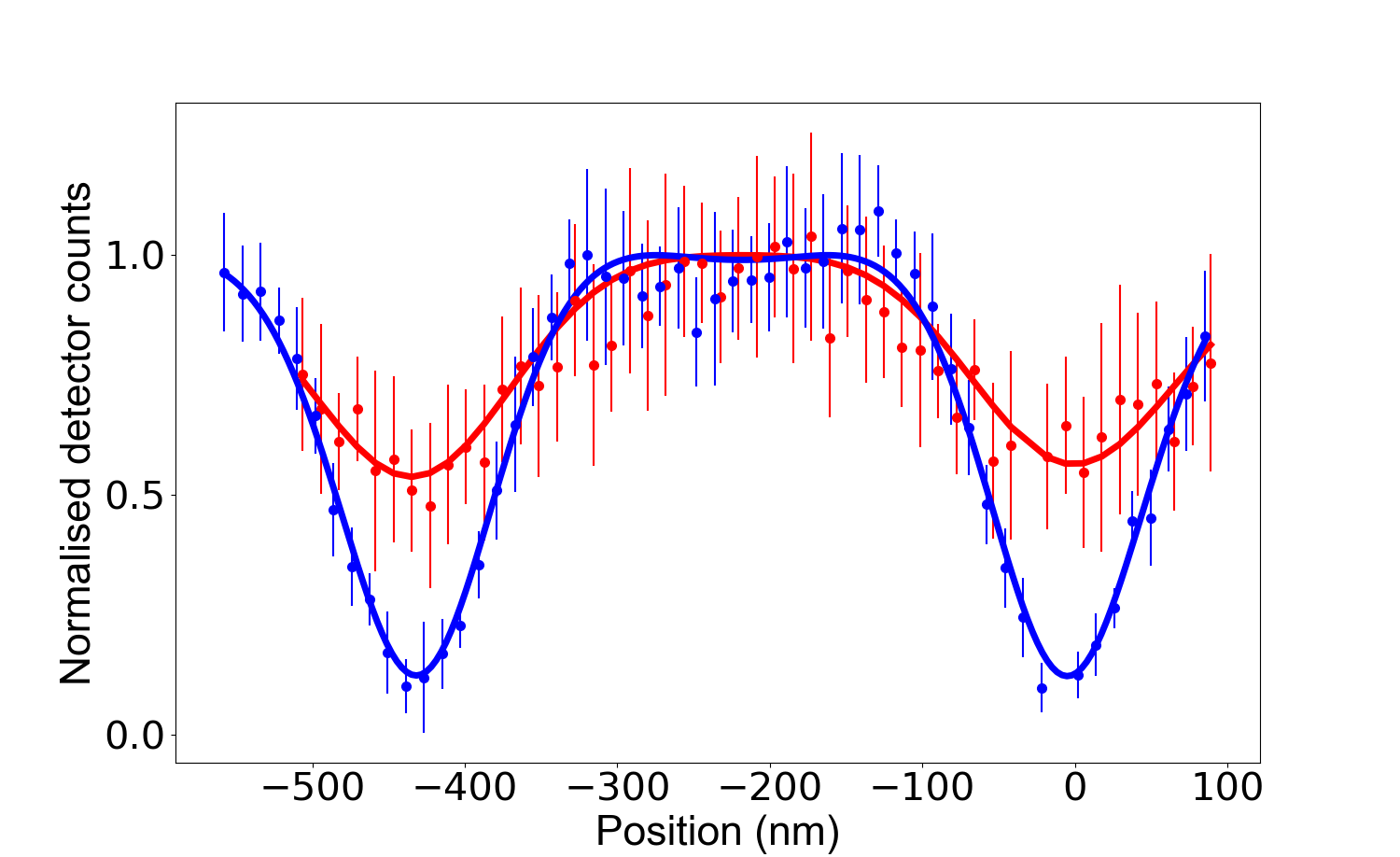}
	\caption{Cavity emission for different ion positions with and without cavity cooling. The solid lines are fits to the measured data. The red line is a trace with Doppler cooling (no cavity cooling) and the blue scan a trace with cavity cooling. The 397 nm cooling laser is detuned -10 MHz from the atomic resonance. The distance between two adjacent nodes is 433 nm.  The error bars represent the statistical mean standard errors.}
	\label{fig_visibilityExample}
\end{figure}

In order to optimise the cavity cooling process, the relevant parameters (cavity detuning, length of periods \RomanNumeralCaps{1} - \RomanNumeralCaps{4} (see Figure \ref{fig_pulse}), cooling beam and probing beam Rabi frequencies) are individually varied. We use the standing wave scans as an indicator for the cavity cooling efficiency. The higher the visibility the more efficient the cavity cooling is. Figure \ref{fig_visibilityExample} shows the fitted final standing wave scan (blue trace), after all the parameters were optimised. The best periods for cooling, shuttling and probing the ion can be seen in Figure \ref{fig_pulse}. The Rabi frequencies are $2\pi\times$14.0 MHz for the cooling 397 nm beam and $2\pi\times$11.8 MHz for the probing 397 nm beam. In addition, for optimal cavity cooling the cavity is blue-detuned 7.0 MHz from the nominal Raman resonance ($\Delta_p$). In the same figure we plot a fitted standing wave scan (red trace) where, instead of cavity cooling, during the same period the ion is Doppler cooled on the \SPh transition, so the cavity does not influence the Doppler cooling process. The 393 nm laser on the \SPh transition has the same angle to the cavity axis as the 397 nm cooling laser and has a similar Doppler cooling efficiency. From Figure \ref{fig_visibilityExample} we can see the improvement that cavity cooling has on the visibility and thus the localisation of the ion. The standing wave pattern measurements are normalised to the maximum emission rate. For Doppler cooling (red trace) the signal to noise ratio was low, because of the delocalisation of the ion due to inefficient cooling, which resulted in big error bars.

 In contrast to \cite{Nature.414.6859}, the pattern is not sinusoidal but consists of wide plateau-like structures at the antinodes. 
 This structure is a result of the strong ion-cavity coupling in the setup. For weak ion-cavity coupling the single photon emission efficiency increases quadratically with coupling \cite{keller2007stable}.
 However, for strong coupling, the emission probability saturates as the maximal efficiency is reached for given pump laser parameters. 
 This is clearly visible in Fig \ref{fig_visibilityExample} (blue trace). If the ion is delocalised due to its thermal motion and hence the effective coherent coupling is reduced, the plateau structure is less pronounced (see red trace in Fig \ref{fig_visibilityExample}).

 To fit the measurement, we first convert the position of the ion to the local coherent coupling using the sinusoidal longitudinal mode pattern. We then convert the local coherent coupling into the expected cavity emission, by numerically solving a master equation of the ion coupled to a bimodal cavity for the given laser parameters. The simulation includes all relevant eight Zeeman sub-levels. We fit the scaling on the x-axis (position) and y-axis (counts), and also keep the localisation due to thermal motion and micromotion minimum offset ($x_0$) as free parameters, where $x_0$ is the ion's distance from the rf centre of the trap. The effect of the thermal motion is well described by a Gaussian position distribution (see \cite{Nature.414.6859}). To approximate the effect of micromotion, we use a Gaussian position distribution with a width given by $\Delta x = \frac{q}{2}x_0$ with $q$ being the trapping q-parameter. Even though this is a coarse approximation and generally underestimates the delocalisation due to micromotion, the agreement with the measurement is good.

\section{Cavity Coupling Strength Measurement}\label{sec_g0}
\begin{figure}[h]
    \centering
	\includegraphics[scale=0.39]{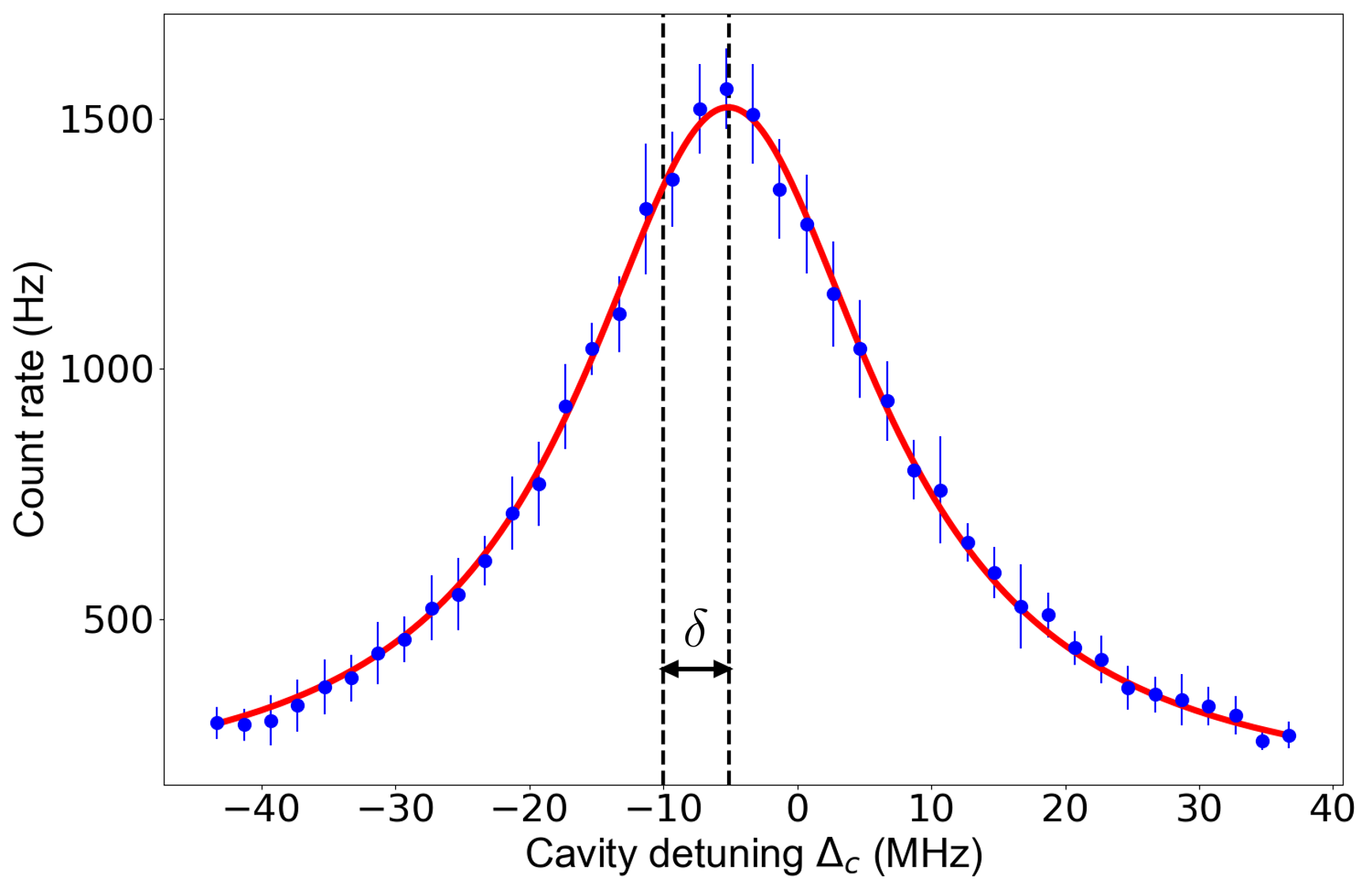}
	\caption{Measured cavity emission vs cavity detuning for a fixed probe detuning of -10 MHz. The red line is a Lorentzian fit to the data. The Raman resonance is shifted by $\delta$ from the expected value.}
	\label{shift}
\end{figure}

 With cavity cooling we can improve the localisation of the ion along the cavity axis, which means that a stronger ion-cavity coupling can be achieved. To quantify the increase of the ion-cavity coupling, we measure it with and without cavity cooling using the method described in \cite{PhysRevLett.124.013602}. The coupling of the ion to the cavity has the effect of shifting the Raman resonance condition by an amount dependent on the coupling. An example of this effect is shown in Figure \ref{shift}. In this case the cavity emission is measured as the cavity is scanned across the Raman resonance while the probe detuning is fixed at $\Delta_p$ = -10 MHz. The Raman resonance is expected at the condition where the cavity detuning is the same as the probe detuning $\Delta_p = \Delta_c$, but instead it has been shifted by an amount $\delta$, as can be seen in Figure \ref{shift}.

\begin{figure}[h]
     \centering
	\includegraphics[scale=0.60]{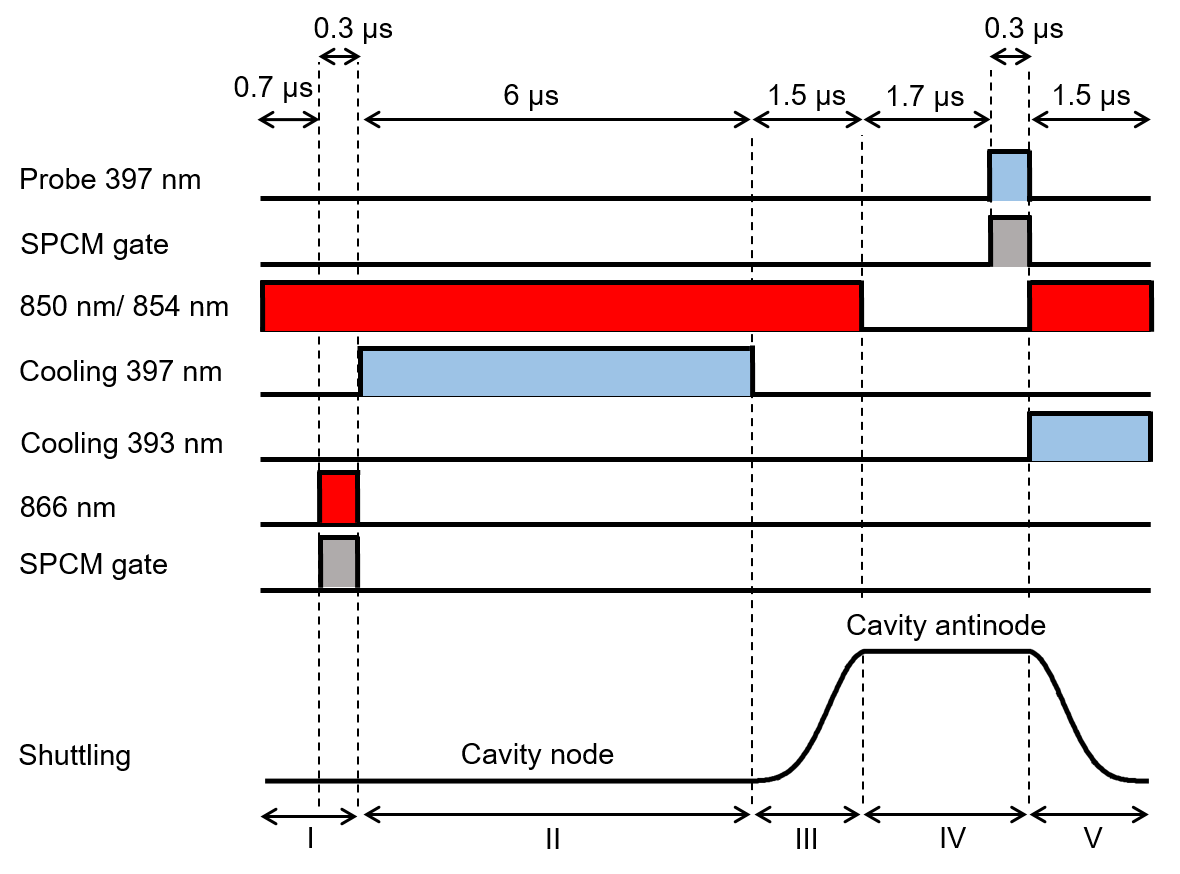}
	\caption{Laser pulses and shuttling voltage pulse during one cycle. (I) State preparation to the $4^{2}\mathrm{S}_{1/2}$ state followed by measurement of the cavity transmission of a laser beam tuned on resonance with the 866 nm atomic transition for reference. Photons are collected by a gated SPCM. (II) Cavity cooling for 6 $\mu s$ with the cavity detuned from the nominal Raman resonance by +7 MHz. (III) Adiabatic shuttling of the ion to the antinode and state preparation. (IV) Generation of a single photon and collection of photons by a gated SPCM. (V) Adiabatic return of the ion to the node with additional Doppler cooling with a 393 nm laser.}
	\label{cavscanpulse}
\end{figure}

 To scan the cavity across the Raman resonance while using the cavity cooling technique we use the pulse sequence displayed in Figure \ref{cavscanpulse}. 
Initially, the ion is prepared in the \Sl state, which decouples the ion from the cavity. The cavity transmission is then measured with an 866 nm laser that is tuned to the atomic transition. 
In this way the reference point of the cavity detuning ($\Delta_c=0$) is determined.
This is followed by a cavity cooling period, where the cavity is blue detuned by 7 MHz from the nominal Raman resonance (this is the optimal cavity detuning for maximising cavity cooling efficiency). The ion is then shuttled adiabatically from the node to the antinode whilst simultaneously preparing the ion in the ground state. This is followed by the single photon generation and detection of the cavity emission via a gated single-photon counting module. Finally, the ion is moved back to the node while simultaneously being Doppler cooled on the \SPh transition. This pulse sequence is repeated as the cavity is scanned across the Raman resonance . As the cavity is scanned during this measurement the cooling beam frequency is varied to keep the detuning of the cavity from the nominal Raman resonance with the cooling beam constant.
 
\begin{figure}[h]
    \centering
	\includegraphics[scale=0.60]{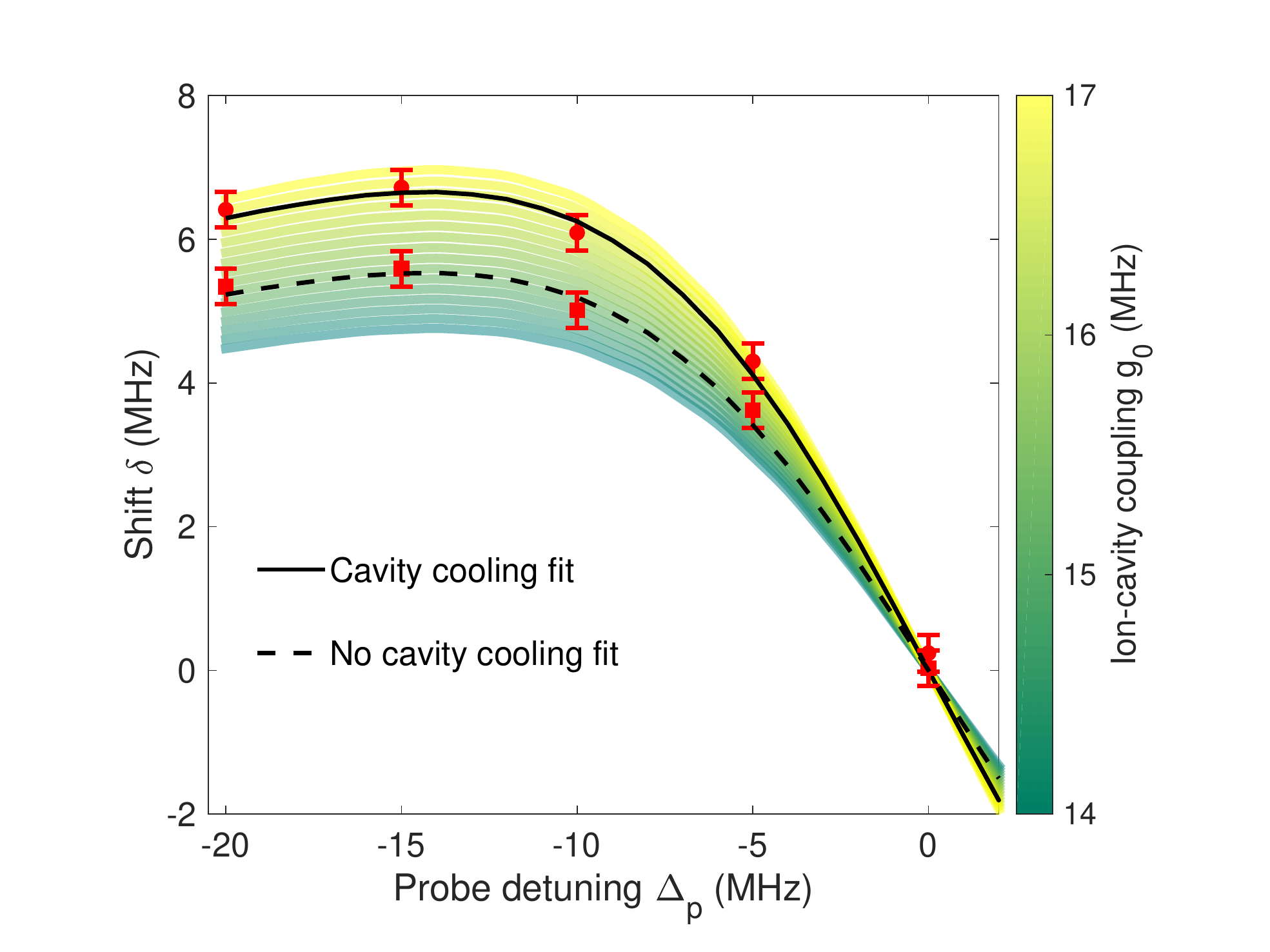}
	\caption{Plots of the shift of the Raman resonance $\delta$ versus the detuning of the probe beam $\Delta_p$. The colour coded map shows the results of the simulation (in units of $2\pi$) that we use to fit the data and extract the coherent coupling. The dashed line shows the fit and the data when there is no cavity cooling and the solid line when there is. The ion-cavity coupling is $2\pi\times (15.2\pm 0.1)$ MHz without cavity cooling and $2\pi\times (16.7\pm 0.1)$ MHz with cavity cooling.}
	\label{fig_g0final}
\end{figure}

 We repeat this Raman spectroscopy for different probe detunings, to measure how $\delta$ changes. Plotting the shift $\delta$ against the probe detuning exhibits a dispersion-like profile. The spectrum is simulated using an 8-level model of \Ca and a bimodal cavity and used to extract the shift $\delta$ for each probe detuning. In this simulation, the only free parameter is the coherent ion-cavity coupling $g_0$. In Figure \ref{fig_g0final} we plot the traces of the simulation for different $g_0$ values. Using this model as a map we fit to our experimental data and extract the coherent ion-cavity coupling. The measured coupling without cavity cooling is $2\pi \times (15.2\pm 0.1)$ MHz (see Figure \ref{fig_g0final}). When including a period of cavity cooling in the sequence (Figure \ref{cavscanpulse}), the coherent ion-cavity coupling increases to $2\pi \times (16.7\pm 0.1)$ MHz. With the expected ion-cavity coupling for a perfectly localised ion of $2\pi\times$17.3 MHz, we calculated that the addition of the cavity cooling step reduces the spread of the ion's position from 110 nm to 55 nm, and the corresponding temperature from 8.5 mK to 2.1 mK. We have increased the measured ion-cavity coupling from 87\% of the theoretically possible ion-cavity coupling with Doppler cooling to 97\% with the introduction of cavity cooling.
 
 While there is a significant improvement in the localisation of the ion through cavity cooling, the effect is significantly smaller than what is expected from a numerical simulation of the cavity cooling process. As the micromotion minimum is 17 nm away from the cavity antinode, the effect of micromotion during measurement of the coupling strength is negligible. However, micromotion may have a significant impact during the cooling phase of the sequence, limiting cooling efficiency, since the micromotion minimum is 200 nm away from the node. The micromotion leads to a reduced effective ion-cavity coupling and subsequently a lowered cooling efficiency. In addition, the driven periodic motion induces spectral sidebands at the trap drive frequency which can further deteriorate the cavity cooling efficiency.

\section{Summary}
In conclusion, we have demonstrated that an optical cavity can be used to significantly improve the localisation of the ion.  By exploiting the strong ion-cavity coupling for cavity cooling instead of circumventing its effect on the laser cooling, we have been able to improve the localisation of the ion in the cavity and thus the effective ion-cavity coupling. While the ion-cavity coupling in our system is limited to $2\pi\times (15.2\pm 0.1)$ MHz by Doppler cooling, we can achieve $2\pi\times (16.7\pm 0.1)$ MHz when employing cavity cooling. This corresponds to 97\% of the maximally achievable coupling in our system. Notably, this is the highest ion-cavity coupling achieved for a single ion in the strong coupling regime.

\newpage
\section{Acknowledgment}

\begin{wrapfigure}{L}{0.10\textwidth}
    \vspace{-13pt}
    \includegraphics[scale=0.018]{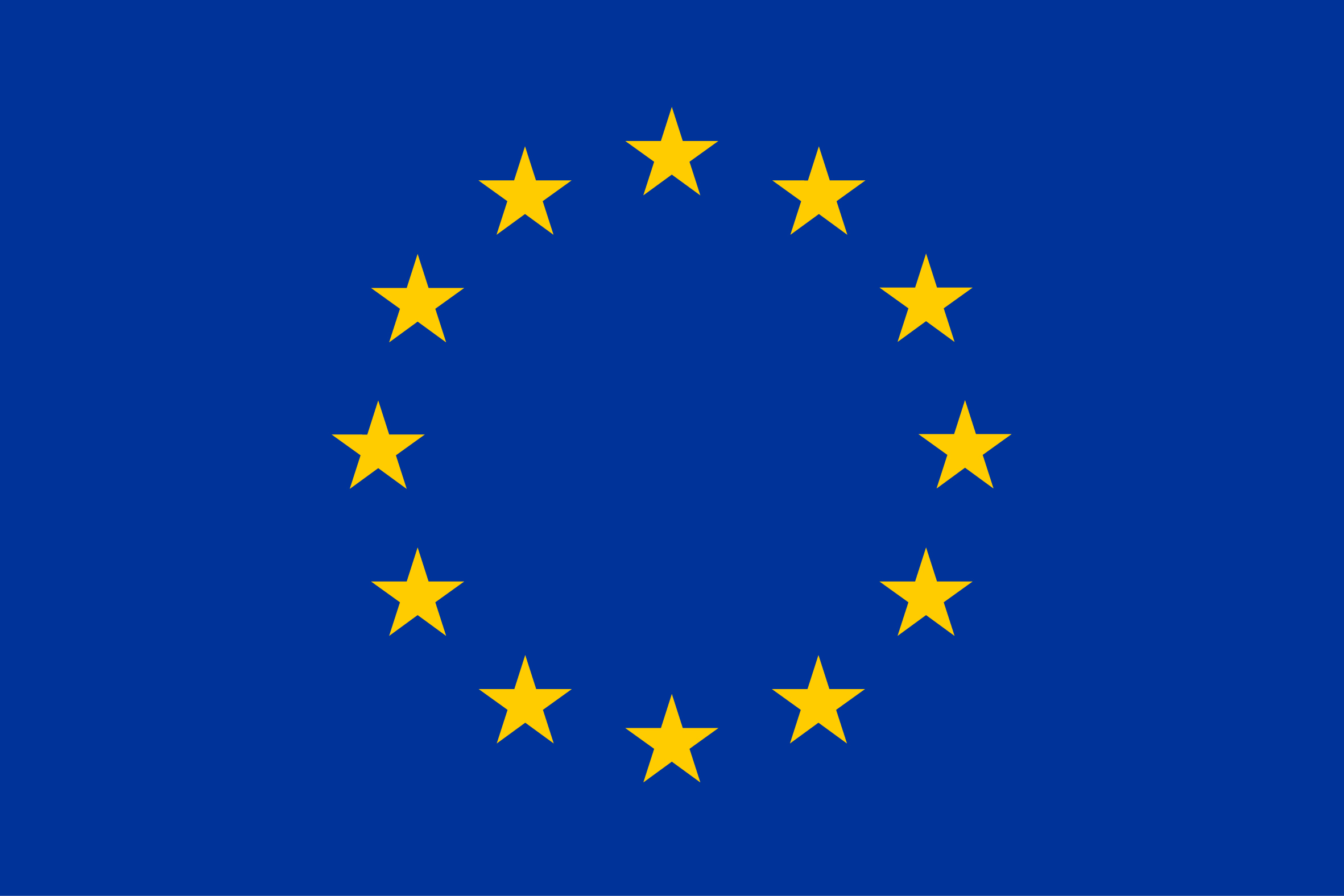}
\end{wrapfigure}
This project has received funding from the European Union’s Horizon 2020 research and innovation program under the Marie Sklodowska-Curie grant agreement No 765075.\newline

We gratefully acknowledge support from EPSRC through the UK Quantum Technology Hub: NQIT - Networked Quantum Information Technologies (EP/M013243/1).



\end{document}